# Toward Turing's A-type Unorganised Machines in an Unconventional Substrate: a Dynamic Representation in Compartmentalised Excitable Chemical Media

Larry Bull, Julian Holley, Ben De Lacy Costello & Andrew Adamatzky

Unconventional Computing Group
University of the West of England, Bristol BS16 1QY, U.K.
`Larry.bull@uwe.ac.uk`

**Abstract**. Turing presented a general representation scheme by which to achieve artificial intelligence – unorganised machines. Significantly, these were a form of discrete dynamical system and yet such representations remain relatively unexplored. Further, at the same time as also suggesting that natural evolution may provide inspiration for search mechanisms to design machines, he noted that mechanisms inspired by the social aspects of learning may prove useful. This paper presents initial results from consideration of using Turing's dynamical representation within an unconventional substrate - networks of Belousov-Zhabotinsky vesicles - designed by an imitation-based, i.e., cultural, approach. Turing's representation scheme is also extended to include a fuller set of Boolean functions at the nodes of the recurrent networks.

## 1  Introduction

In 1948 Alan Turing produced an internal paper where he presented a formalism he termed "unorganised machines" by which to represent intelligence within computers (eventually published in [39]). These consisted of two main types: A-type unorganised machines, which were composed of two-input NAND gates connected into disorganised networks (Figure 1); and, B-type unorganised machines which included an extra triplet of NAND gates on the arcs between the NAND gates of A-type machines by which to affect their behaviour in a supervised learning-like scheme. In both cases, each NAND gate node updates in parallel on a discrete time step with the output from each node arriving at the input of the node(s) on each connection for the next time step. The structure of unorganised machines is therefore very much like a simple artificial neural network with recurrent connections and hence it is perhaps surprising that Turing made no reference to McCulloch and Pitts' [29] prior seminal paper on networks of binary-thresholded nodes. However, Turing's scheme extended McCulloch and Pitts' work in that he also considered the training of such networks with his B-type architecture. This has led to their also being known as "Turing's connectionism" (e.g., [10]). Moreover, as Teuscher [36] has highlighted, Turing's unorganised machines are (discrete) nonlinear dynamical systems and therefore have the potential to exhibit complex behaviour despite their construction from simple elements. The current work aims to explore the use of Boolean dynamic system representations within networks of small lipid-coated vesicles. The excitable chemical Belousov-Zhabotinsky (BZ) [42] medium is packaged into the vesicles which form the simple/elementary components of a liquid information processing system. The vesicles communicate through chemical "signals" as excitation propagates from vesicle to vesicle. Initial experimental implementations which use micro-fluidics to control vesicle placement have recently been reported [25].

This paper begins by considering implementation of the basic two-input NAND gates using the vesicles and then how to design networks of vesicles to perform a given computation. In particular, a form of collision-based computing (e.g., [1]) is used, along with imitation programming (IP) [8], which was also inspired by Turing's 1948 paper, specifically the comment that "*Further research into intelligence of machinery will probably be very greatly concerned with 'searches' .... [an example] form of search is what I should like to call the 'cultural search' ... the search for new techniques must be regarded as carried out by the human community as a whole*" [39]. Kauffman [22] introduced a form of dynamical Boolean network which uses any possible Boolean function at each node – random Boolean networks (RBN). The use of other well-known Boolean functions within the networks is subsequently explored here, again through collision-based computing. Performance from the extension to more realistic signal propagation times within the networks is also explored.

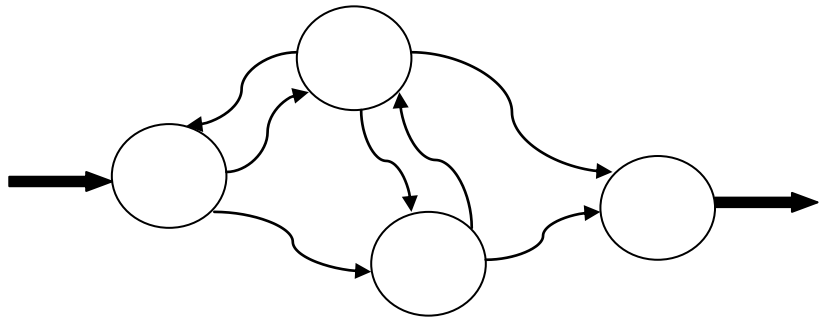

**Fig. 1.** Simple example A-type unorganised machine consisting of four two-input NAND gate nodes ($N$=4), with one input (node 1) and one output (node 4) as indicated by the bold arrows.

## 2   Background

*"The machine is made up from a rather large number N of similar units. Each unit has two input terminals, and has an output terminal which can be connected to input terminals of (0 or more) other units. We may imagine that for each integer r, $1 \leq r \leq N$, two numbers i(r) and j(r) are chosen at random from 1..N and that we connect the inputs of unit r to the outputs of units i(r) and j(r). All of the units are connected to a central synchronising unit from which synchronising pulses are emitted at more or less equal intervals of time. The times when these pulses arrive will be called 'moments'. Each unit is capable of having two states at each moment. These states may be called 0 and 1. The state is determined by the rule that the states of the units from the input leads come are to be taken at the previous moment, multiplied together and then subtracted from 1"* [39].

A-type unorganised machines have a finite number of possible states and they are deterministic, hence such networks eventually fall into a basin of attraction. Turing was aware that his A-type unorganised machines would have periodic behaviour and he stated that since they represent "*about the simplest model of a nervous system with a random arrangement of neurons*" it would be "*of very great interest to find out something about their behaviour*" [39]. Figure 2 shows the fraction of nodes which change state per update cycle for 100 randomly created networks, each started from a random initial configuration, for various numbers of nodes *N*. As can be seen, the time taken to equilibrium is typically around 15 cycles, with all nodes in the larger case changing state on each cycle thereafter, i.e., oscillating (see also [36]). For the smaller networks, some nodes remain unchanging at equilibrium on average; with smaller networks, the probability of nodes being isolated is sufficient that the basin of attraction contains a degree of node stasis. However, there is significant variance in behaviour.

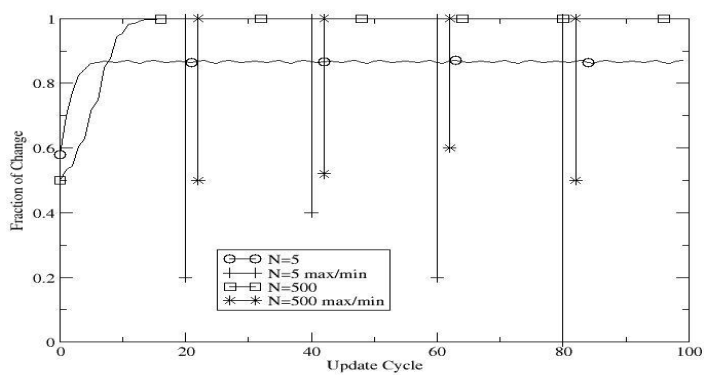

**Fig. 2.** Showing the average fraction of two-input NAND gate nodes which change state per update cycle of random A-type unorganised machines with various numbers of nodes *N*. Error bars show max. and min. values from 100 trials.

Turing [39] envisaged his A-type unorganised machines being used such that they " *... are allowed to continue in their own way for indefinite periods without interference from outside*" and went on to suggest that one way to use them for computation would be to exploit how the application of external inputs would alter the (dynamic) behaviour of the machine. This can be interpreted as his suggesting individual attractors are used to represent distinct (discrete) machine states and the movement between different attractors as a result of different inputs a way to perform computation. Note this hints at some of the ideas later put forward by Ashby [6] on brains as dynamic systems.

Teuscher [36] used a genetic algorithm (GA) [18] to design A-type unorganised machines for bitstream regeneration tasks and simple pattern classification. Bull [8] used IP to design simple logic circuits, such as multiplexers, from them. Here the unorganised machine had an external input applied, was then updated for some number of cycles, e.g., sufficient for an attractor to be typically reached, and then the state of one or more nodes was used to represent the output. More generally, it is well-established that discrete dynamical systems can be robust to faults, can compute, can exhibit memory, etc. (e.g., see [23][41]).

Given their relative architectural simplicity but potential for complex behaviour, A-type unorganised machines appear to be a good candidate (dynamic) representation to use with novel computing substrates. Their use for a chemical computing system is considered here. It can be noted that Turing (e.g., [40]) was also interested in chemical reaction-diffusion systems, for pattern formation not computation.

## 3   Chemical Computing

Excitable and oscillating chemical systems have been used to solve a number of computational tasks such as implementing logical circuits [34], image processing [26], shortest path problems [33] and memory [31]. In addition chemical diodes [5], coincidence detectors [15] and transformers where a periodic input signal of waves may be modulated by the barrier into a complex output signal depending on the gap width and frequency of the input [32] have all been demonstrated experimentally.

A number of experimental and theoretical constructs utilising networks of chemical reactions to implement computation have been described. These chemical systems act as simple models for networks of coupled oscillators such as neurons, circadian pacemakers and other biological systems [24]. Ross and co-workers [16] produced a theoretical construct suggesting the use of "chemical" reactor systems coupled by mass flow for implementing logic gates neural networks and finite-state machines. In further work Hjelmfelt et al. [17] simulated a pattern recognition device constructed from large networks of mass-coupled chemical reactors containing a bistable iodate-arsenous acid reaction. They encoded arbitrary patterns of low and high iodide concentrations in the network of 36 coupled reactors. When the network is initialized with a pattern similar to the encoded one then errors in the initial pattern are corrected bringing about the regeneration of the stored pattern. However, if the pattern is not similar then the network evolves to a homogenous state signalling non-recognition.

In related experimental work Laplante et al. [27] used a network of eight bistable mass coupled chemical reactors (via 16 tubes) to implement pattern recognition operations. They demonstrated experimentally that stored patterns of high and low iodine concentrations could be recalled (stable output state) if similar patterns were used as input data to the programmed network. This highlights how a programmable parallel processor could be constructed from coupled chemical reactors. This described chemical system has many properties similar to parallel neural networks. In other work Lebender and Schneider [28] described methods of constructing logical gates using a series of flow rate coupled continuous flow stirred tank reactors (CSTR) containing a bistable nonlinear chemical reaction. The minimal bromate reaction involves the oxidation of cerium(III) ($Ce^{3+}$) ions by bromate in the presence of bromide and sulphuric acid. In the reaction the $Ce^{4+}$ concentration state is considered as "0" "false" ("1""true") if a given steady state is within 10% of the minimal (maximal) value. The reactors were flow rate coupled according to rules given by a feedforward neural network run using a PC. The experiment is started by feeding in two "true" states to the input reactors and then switching the flow rates to generate "true"-"false", "false"-"true" and "false"-"false". In this three coupled reactor system the AND (output "true" if inputs are both high $Ce^{4+}$, "true"), OR (output "true" if one of the inputs is "true"), NAND (output "true" if one of the inputs is "false") and NOR gates (output "true" if both of the inputs are "false") could be realised. However to construct XOR and XNOR gates two additional reactors (a hidden layer) were required. These composite gates are solved by interlinking AND and OR gates and their negations. In their work coupling was implemented by computer but they suggested that true chemical computing of some Boolean functions may be achieved by using the outflows of reactors as the inflows to other reactors, i.e., serial mass coupling.

As yet no large scale experimental network implementations have been undertaken mainly due to the complexity of analysing and controlling many reactors. That said there have been many experimental studies carried out involving coupled oscillating and bistable systems (e.g., see [35][11][7][21]). The reactions are coupled together either physically by diffusion or an electrical connection or chemically, by having two oscillators that share a common chemical species. The effects observed include multistability, synchronisation, in-phase and out of phase entrainment, amplitude or "oscillator death", the cessation of oscillation in two coupled oscillating systems, or the converse, "rhythmogenesis", in which coupling two systems at steady state causes them to start oscillating [13].

Vesicles formed from droplets of BZ medium (Figure 3), typically just a few millimetres in diameter, exhibit many properties which may be considered as rudimentary for possible future molecular processing systems: signal transmission, self-repair, signal gain, self-organisation, etc. Their potential use for computation has begun to be explored through collision-based schemes (e.g., [3][4][19][20]). This paper considers their use within a dynamic representation using a collision-based scheme.

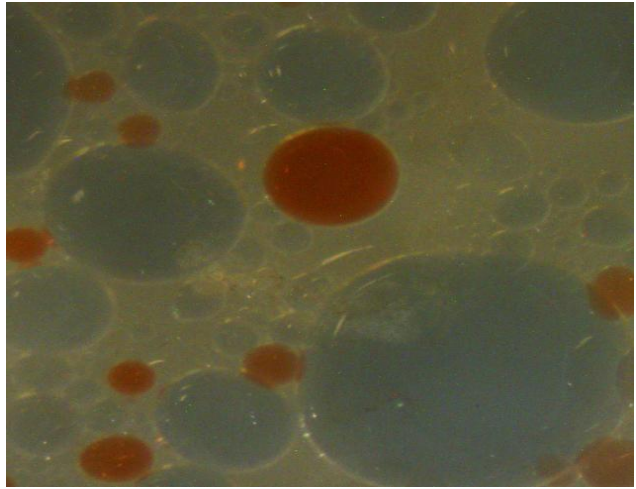

**Fig. 3.** Showing the BZ droplet vesicles.

Collision-based computing exploits the interaction of moving elements and their mutual effects upon each other's movement wherein the presence or absence of elements at a given point in space and time can be interpreted as computation (e.g., see [2] for chemical systems). Collision-based computing is here envisaged within recurrent networks of BZ vesicles, i.e., based upon the movement and interaction of waves of excitation within and across vesicle membranes. For example, to implement a two-input NAND gate, consider the case shown in Figure 4: when either input is applied, as a stream of waves of excitation, no waves are seen at the output location in the top vesicle - only when two waves coincide is a wave subsequently seen at the output location giving logical AND. A NOT gate can be constructed through the disruption of a constant Truth input in another vesicle, as shown.

A-type unorganised machines can therefore be envisaged within networks of BZ vesicles using the three-vesicle construct for the NAND gate nodes, together with chains of vesicles to form the connections between them. Creation of such chains is reported in the initial experimentation with micro-fluidics noted above [25]. As also noted above, it has recently been shown that IP is an effective design approach with the dynamic representation.

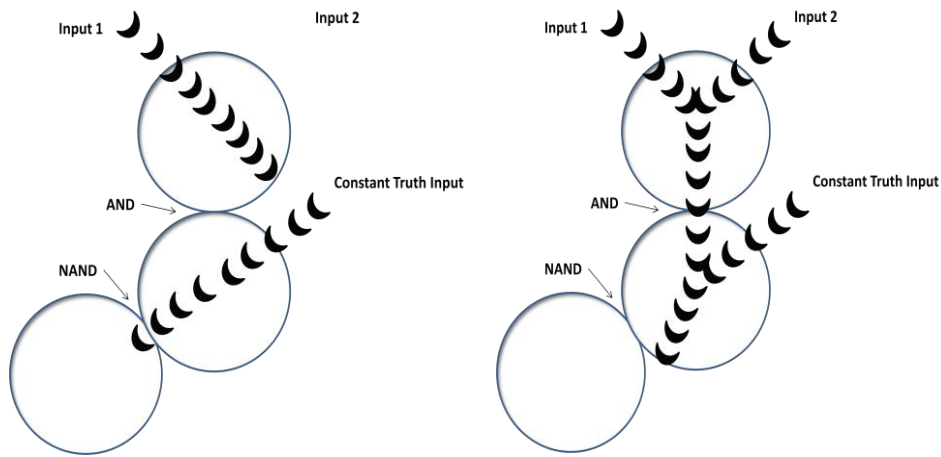

**Fig. 4.** Showing the construction of a two-input NAND gate under a collision-based scheme using three BZ vesicles. The cases of inputs True-False (left) and True-True (right) are shown. Note the existence of an AND gate also.

## 4   Imitation Programming

The basic principle of imitation programming is that individuals alter themselves based upon another individual(s), typically with some error in the process. Individuals are not replaced with the descendants of other individuals as in evolutionary search; individuals persist through time, altering their solutions via imitation. Thus imitation may be cast as a directed stochastic search process, thereby combining aspects of both recombination and mutation used in evolutionary computation:

```
BEGIN
INITIALISE population with random candidate solutions
EVALUATE each candidate
REPEAT UNTIL (TERMINATION CONDITION) DO
    FOR each candidate solution DO
          SELECT candidate(s) to imitate
          CHOOSE component(s) to imitate
          COPY the chosen component(s) with ERROR
          EVALUATE new solution
          REPLACE IF (UPDATE CONDITION) candidate with new solution
    OD
OD
END
```

For A-type design, IP utilizes a variable-length representation of pairs of integers defining node inputs, each with an accompanying single bit defining the node's start state. There are three imitation operators - copy a node connection, copy a node start state, and change size through copying. In this paper, each operator can occur with or without error, with equal probability, such that an individual performs one of the six during the imitation process as follows:

To copy a node connection, a randomly chosen node has one of its randomly chosen connections set to the same value as the corresponding node and its same connection in the individual it is imitating. When an error occurs, the connection is set to the next or previous node (equal probability, bounded by solution size). Imitation can also copy the start state for a randomly chosen node from the corresponding node, or do it with error (bit flip here). Size is altered by adding or deleting nodes and depends upon whether the two individuals are the same size. If the individual being imitated is larger than the copier, the connections and node start state of the first extra node are copied to the imitator, a randomly chosen node being connected to it. If the individual being imitated is smaller than the copied, the last added node is cut from the imitator and all connections to it re-assigned. If the two individuals are the same size, either event can occur (with equal probability). Node addition adds a randomly chosen node from the individual being imitated onto the end of the copier and it is randomly connected into the network. The operation can also occur with errors such that copied connections are either incremented or decremented. For a problem with a given number of binary inputs $I$ and a given number of binary outputs $O$, the node deletion operator has no effect if the parent consists of only $O + I + 2$ nodes. The extra two inputs are constant True and False lines. Similarly, there is a maximum size (100) defined beyond which the growth operator has no effect.

In this paper, each individual in the population $P$ creates one variant of itself and it is adopted if better per iteration. In the case of ties, the solution with the fewest number of nodes is kept to reduce size, otherwise the decision is random. The individual to imitate is chosen using a roulette-wheel scheme based on proportional solution utility, i.e., the traditional reproduction selection scheme used in GAs. Other forms of updating, imitation processes, and imitation selection are, of course, possible [8]. In this form IP may be seen as combining ideas from memetics [12] with Evolutionary Programming [14]. It can be noted GAs have previously been used to design chemical computing systems in various ways (e.g., [9][37][38]).

## 5   A-type Experimentation

In the following, three well-known logic problems are used to begin to explore the characteristics and capabilities of the general approach. The multiplexer task is used since they can be used to build many other logic circuits, including larger multiplexers. These Boolean functions are defined for binary strings of length $l = k + 2^k$ under which the $k$ bits index into the remaining $2^k$ bits, returning the value of the indexed bit. Hence the multiplexer has multiple inputs and a single output. The demultiplexer and adders have multiple inputs and multiple outputs. As such, simple examples of each are also used here. A simple sequential logic task is also used here - the JK latch. In all cases, the correct response to a given input results in a quality increment of 1, with all possible binary inputs being presented per solution evaluation. Upon each presentation of an input, each node in an unorganised machine has its state set to its specified start state. The input is applied to the first connection of each corresponding $I$ input node. The A-type is then executed for 15 cycles. The value on the output node(s) is then taken as the response. All results presented are the average of 20 runs, with $P=20$. Experience found giving initial random solutions $N=O+I+2+30$ nodes was useful across all the problems explored here, i.e., with the other parameter/algorithmic settings.

Figure 5 shows the performance of IP to design A-type unorganised machines on $k=2$ versions of the four tasks: the 6-bit multiplexer (opt. 64), 2-bit adder (opt. 16), 6-bit demultiplexer (opt. 8) and 2-input JK latch (opt. 4). As can be seen, optimal performance is reached in all cases, well within the allowed time, and that the solution sizes are adjusted to the given task. That is, discrete dynamical circuits capable of the given logic functions have been designed. As discussed elsewhere [8], the relative robustness of such circuits to faults, their energy usage, etc. remains to be explored.

However, to begin to consider implementing such designs within BZ vesicles, the time taken for signal propagation between NAND gate nodes needs to included. That is, in Figure 5, as in all previous work with such dynamic representations, any changes in node state are immediately conveyed to any other connected nodes since a traditional computational substrate is assumed. Within the vesicles, changes in NAND gate node state will propagate through chains and hence there will be a time delay proportional to the distance between nodes.

Figure 6 shows results for the same experiments and parameters as before but with a form of time delay added to begin to consider the physical implementation in an elementary way. Here NAND gate node states take the same number of update cycles to propagate between nodes as the absolute difference in node number. For example, the state of node 11 at time $t$ would take 8 update cycles to reach node 3. Hence at update cycle $t+8$, node 3 would use the state of node 11 as at time $t$ as one of its inputs. The number of overall update cycles for the A-types was increased to 50 to help facilitate signal passing across the network.

As Figure 6 shows, it takes longer to reach optimal solutions (T-test, $p<0.05$) and they are perhaps surprisingly smaller (T-test, $p<0.05$, except JK Latch) than before, but suitable dynamic designs are again found in the allotted time, except for the adder which takes longer to reach optimality.

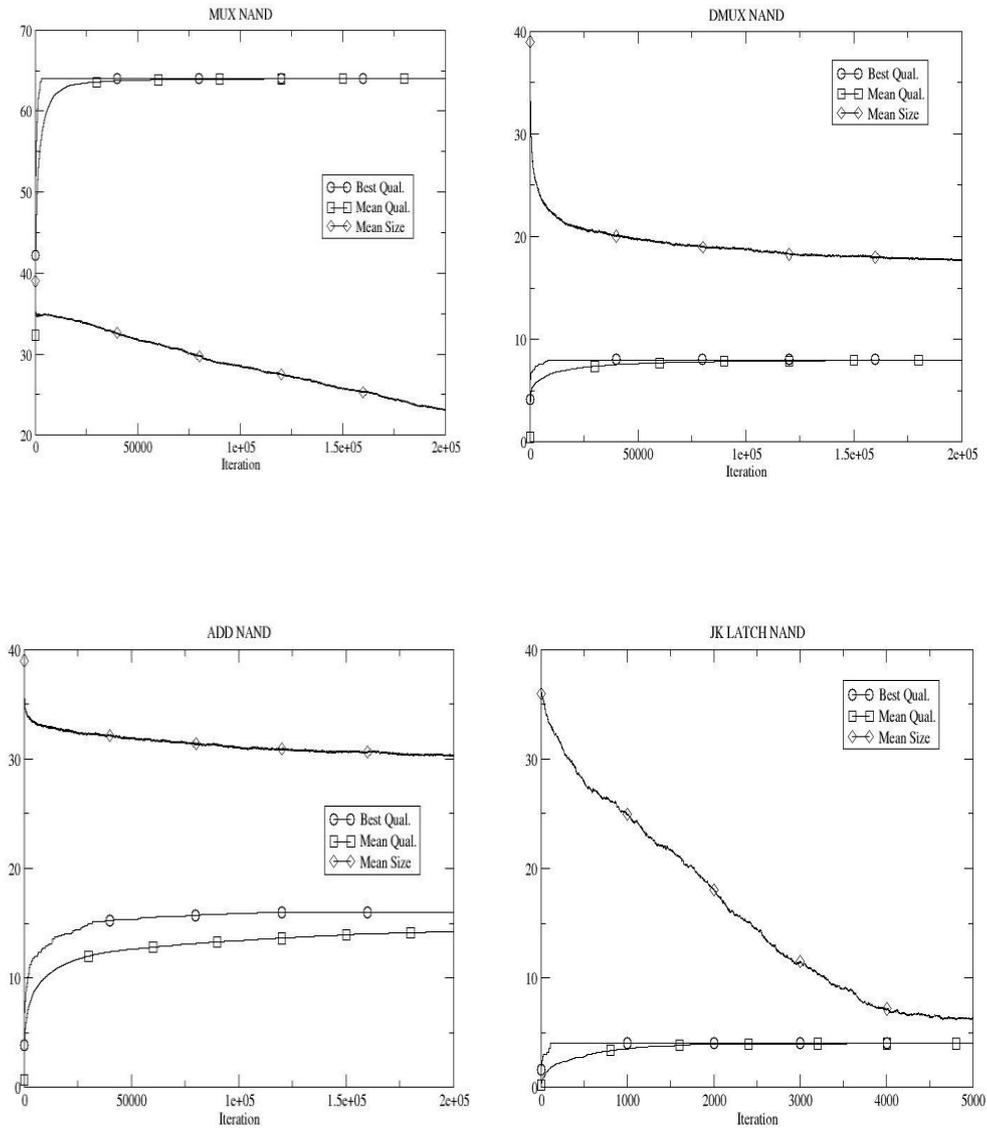

**Fig. 5.** Showing the performance of IP in designing A-type unorganised machines for the three combinatorial and single sequential logic tasks.

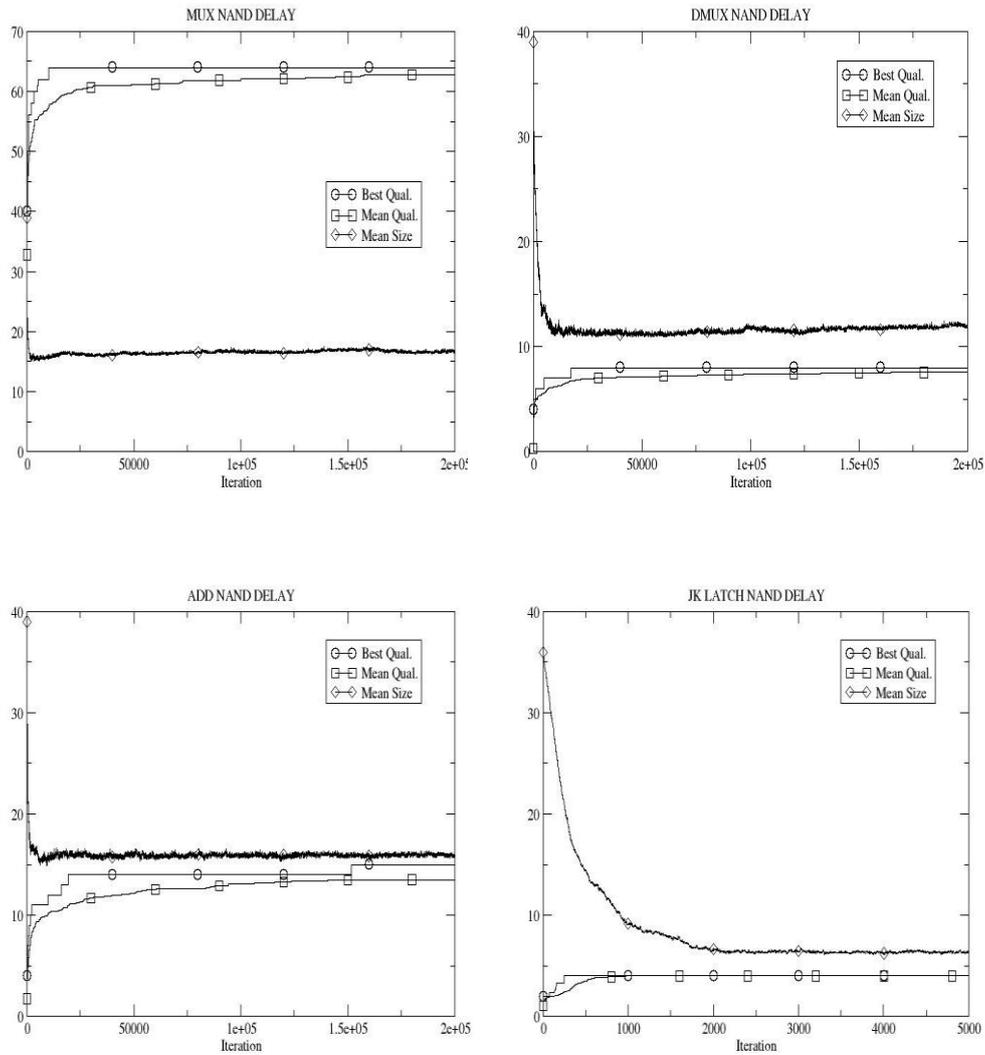

**Fig. 6.** Showing the performance of IP in designing A-type unorganised machines for the four logic tasks with signal propagation times added.

## 6　RBN Experimentation

Random Boolean networks were originally introduced to explore aspects of biological genetic regulatory networks. Since then they have been used as a tool in a wide range of areas such as self-organisation (e.g., [23]) and computation (e.g., [30]). An RBN typically consists of a network of $N$ nodes, each performing a Boolean function with $K$ inputs from other nodes in the network, all updating synchronously. As such, RBN may be viewed as a generalization of Turing's A-type unorganised machines which used only the NAND Boolean function with $K=2$. As noted above, Turing's paper was not published until 1968 so it is perhaps not too surprising that Kauffman did not originally discuss his work - although no connection has been made subsequently either, except in [36].

It is well-established that the value of $K$ affects the emergent behaviour of RBN wherein attractors typically contain an increasing number of states with increasing $K$. Three phases of behaviour are suggested: ordered when $K=1$, with attractors consisting of one or a few states; chaotic when $K>2$, with a very large numbers of states per attractor; and, a critical regime when $K=2$, where similar states lie on trajectories that tend to neither diverge nor converge and 5-15% of nodes change state per attractor cycle (see [23] for discussions of this critical regime, e.g., with respect to perturbations). Analytical methods have been presented by which to determine the typical time taken to reach a basin of attraction and the number of states within such basins for a given degree of connectivity $K$ (again, see [23]).

The previous A-type scenario has been extended to include other well-known Boolean functions through collision-based schemes. Figures 7 and 8 show how two-input OR, NOR, XOR and XNOR can all be achieved using vesicles.

Figure 9 shows how performance is not typically improved in any case considered (T-test, $p<0.05$) with the extra Boolean functionality added - AND, NAND, OR, NOR, XOR, XNOR – and hence Turing's simpler scheme appears to represent a potentially useful approach for implementation with the vesicles.

## 7　Conclusions

Over sixty years ago, Alan Turing presented a simple representation scheme for machine intelligence – a discrete dynamical system network of two-input NAND gates. Since then only a few other explorations of these unorganized machines are known. As noted above, it has long been argued that dynamic representations provide numerous useful features, such as an inherent robustness to faults and memory capabilities by exploiting the structure of their basins of attraction. For example, unique attractors can be assigned to individual system states/outputs and the map of internal states to those attractors can be constructed such that multiple paths of similar states lead to the same attractor. In this way, some variance in the actual path taken through states can be varied, e.g., due to errors, with the system still responding appropriately. Turing appears to have been thinking along these lines also.

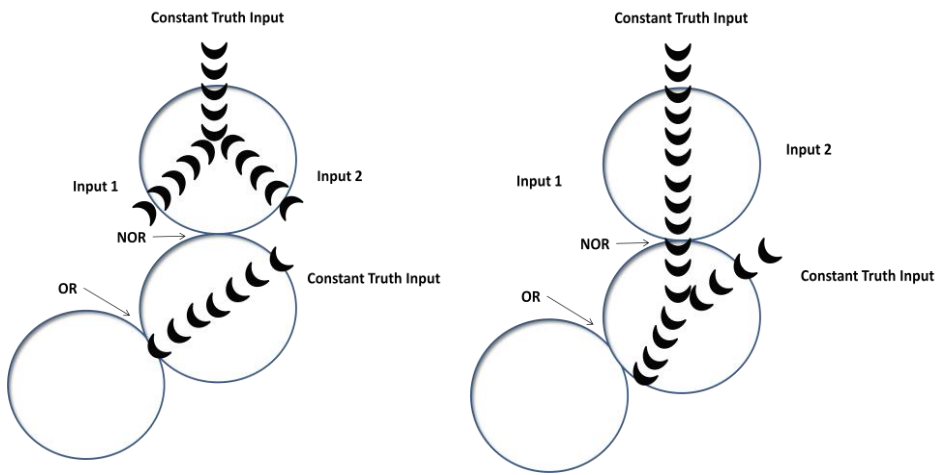

**Fig. 7.** Showing the construction of a two-input OR and NOR gates under a collision-based scheme using three BZ vesicles. The cases of inputs False-False (left) and True-True (right) are shown.

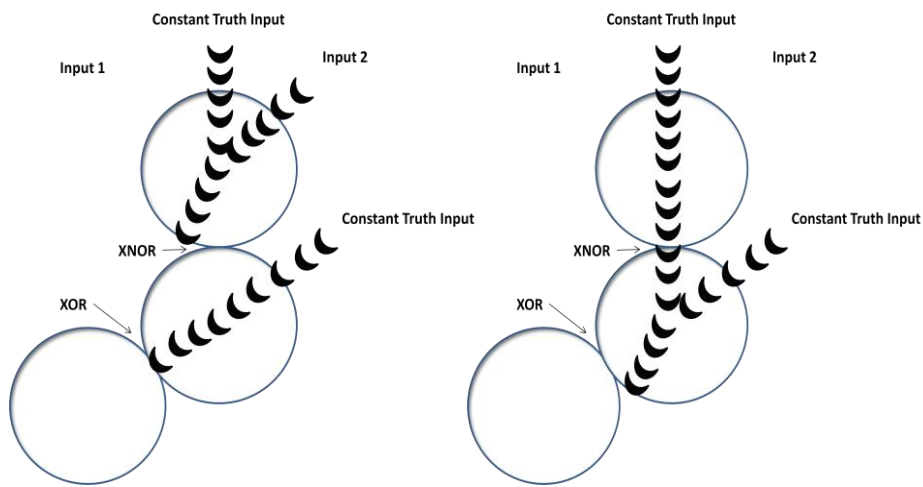

**Fig. 8.** Showing the construction of a two-input XOR and XNOR gates under a collision-based scheme using three BZ vesicles. The cases of inputs False-False (left) and False-True (right) are shown.

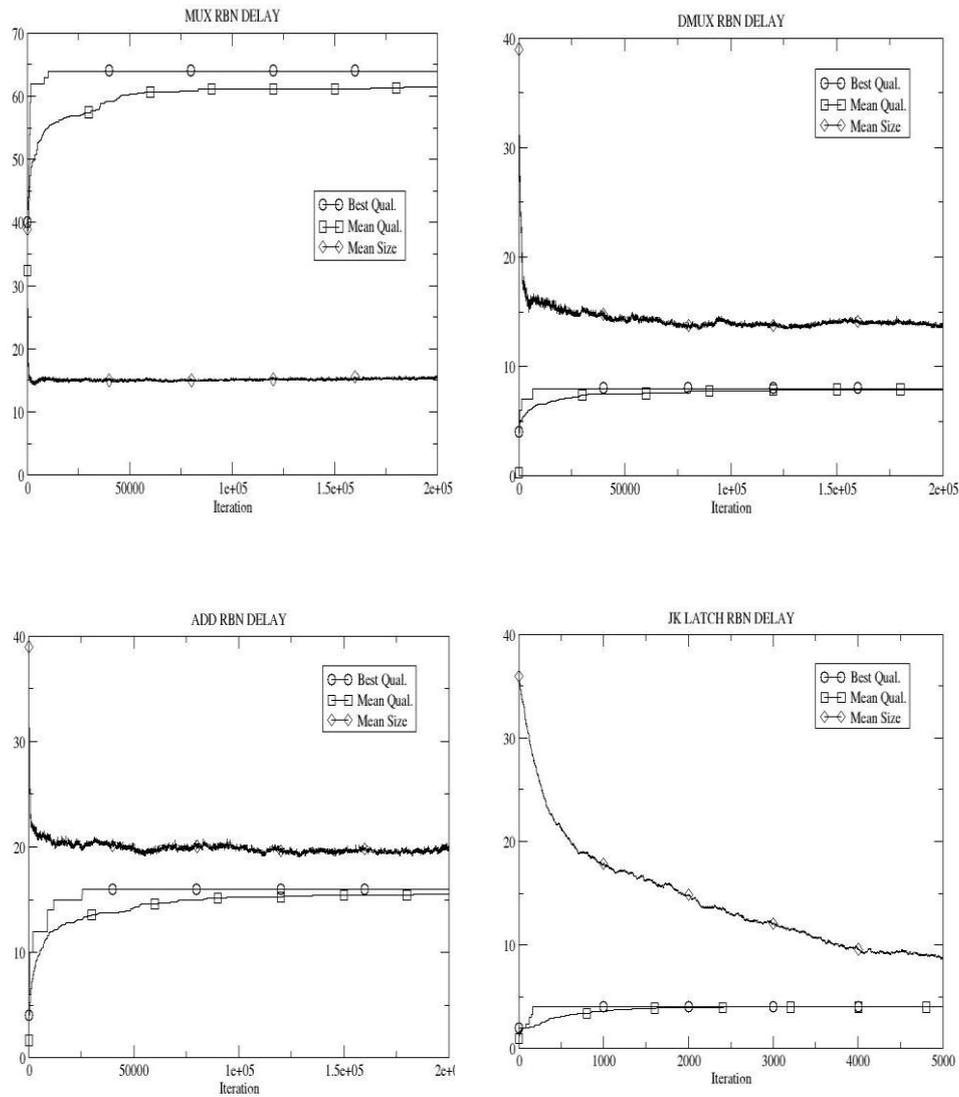

**Fig. 9.** Showing the performance of IP in designing RBN for the four logic tasks with signal propagation times.

Given the relative simplicity of A-types but their potential for complex behaviour, this paper suggests they may provide a useful representation scheme for unconventional computing substrates. Unconventional computing aims to go beyond traditional architectures and formalisms, much of which is based upon Turing's work on computability, by exploiting the inherent properties of systems to perform computation. A number of experimental systems have been presented in biological, chemical and physical media. Where NAND gate function can be realised, whilst also leaving open the potential utilisation of other aspects of the chosen medium, A-types could be explored. In particular, a substrate of BZ vesicles recently presented as a step towards molecular information processing, e.g., for future smart drugs, was considered and a form of two-input NAND gate designed for it through collision-based computing.

It was then shown how a number of well-known benchmark logic circuits can be designed from A-type unorganised machines using an approach inspired by a comment from Turing on cultural search. Further consideration of the physical implementation within networks of BZ vesicles meant that signal propagation times were also included into the A-types. Results indicate that the design process was slowed relatively but still effective. Extending the NAND gate functionality to include other well-known Boolean logic within the networks showed no improved performance in the more realistic case. Current work is increasing the level of detail of the simulated chemical system both in terms of the vesicle structure and of the BZ reaction therein.

**Acknowledgement**

The research was supported by the NEUNEU project sponsored by the European Community within FP7-ICT-2009-4 ICT-4-8.3 - FET Proactive 3: Bio-chemistry-based Information Technology (CHEM-IT) program